\documentclass[journal=jpclcd,manuscript=letter]{achemso}

\usepackage{siunitx}
\usepackage{amssymb}

\author{Zhuo Yang}
\altaffiliation{These authors contributed equally to the work}
\affiliation{Laboratoire National des Champs Magn\'etiques Intenses, UPR 3228, CNRS-UGA-UPS-INSA, Grenoble and Toulouse, France}

\author{Alessandro Surrente}\altaffiliation{These authors contributed equally to the work}
\affiliation{Laboratoire National des Champs Magn\'etiques Intenses, UPR 3228, CNRS-UGA-UPS-INSA, Grenoble and Toulouse, France}

\author{Krzysztof Galkowski}
\affiliation{Laboratoire National des Champs Magn\'etiques Intenses, UPR 3228, CNRS-UGA-UPS-INSA, Grenoble and Toulouse, France}\alsoaffiliation{Institute of Experimental Physics, Faculty of
Physics, University of Warsaw - Pasteura 5, 02-093 Warsaw, Poland}

\author{Nicolas Bruyant}
\affiliation{Laboratoire National des Champs Magn\'etiques Intenses, UPR 3228, CNRS-UGA-UPS-INSA, Grenoble and Toulouse, France}

\author{Duncan K.\ Maude}
\affiliation{Laboratoire National des Champs Magn\'etiques Intenses, UPR 3228, CNRS-UGA-UPS-INSA, Grenoble and Toulouse, France}

\author{Amir Abbas Haghighirad}
\affiliation{University of Oxford, Clarendon Laboratory, Parks Road, Oxford, OX1 3PU, United Kingdom}

\author{Henry J.\ Snaith}
\affiliation{University of Oxford, Clarendon Laboratory, Parks Road, Oxford, OX1 3PU, United Kingdom}

\author{Paulina Plochocka}\email{paulina.plochocka@lncmi.cnrs.fr}
\affiliation{Laboratoire National des Champs Magn\'etiques Intenses, UPR 3228, CNRS-UGA-UPS-INSA, Grenoble and Toulouse, France}
\email{paulina.plochocka@lncmi.cnrs.fr}
\phone{+33 (0) 562 17 28 62}

\author{Robin J.\ Nicholas}\email{r.nicholas@physics.ox.ac.uk}
\affiliation{University of Oxford, Clarendon Laboratory, Parks Road, Oxford, OX1 3PU, United Kingdom}
\email{r.nicholas@physics.ox.ac.uk}
\phone{+44-1865-272250 }

\title{Unraveling the Exciton Binding Energy and the Dielectric Constant in Single Crystal Methylammonium Lead Tri-Iodide Perovskite}

\begin{document}

\begin{tocentry}

    \includegraphics[width= 5.0cm]{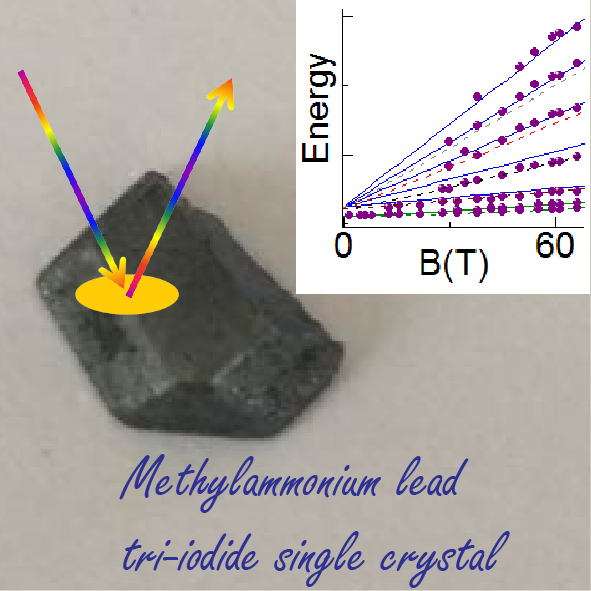}

\end{tocentry}

\begin{abstract}
We have accurately determined the exciton binding energy and reduced mass of single crystals of methylammonium lead tri-iodide
using magneto-reflectivity at very high magnetic fields. The single crystal has excellent optical properties with a narrow line
width of $\sim 3$\,meV for the excitonic transitions and a 2s transition which is clearly visible even at zero magnetic field.
The exciton binding energy of $16 \pm 2$\,meV in the low temperature orthorhombic phase is almost identical to the value found in
polycrystalline samples, crucially ruling out any possibility that the exciton binding energy depends on the grain size. In the
room temperature tetragonal phase, an upper limit for the exciton binding energy of $12 \pm 4$\,meV is estimated from the
evolution of 1s-2s splitting at high magnetic field.
\end{abstract}

Solar cells based on hybrid organic-inorganic perovskites have
demonstrated an exceptionally rapid increase in the photovoltaic
energy conversion efficiency, which has recently reached record
values of more than
20\%\cite{yang2015high,saliba2016cesium,li2016vacuum,saliba2016incorporation}. These materials crystallize in the form ABX$_3$, where A is an
organic ammonium cation (typically methylammonium, MA, or
formamidinium, FA), B is Pb or Sn and X is a halide, with
methylammonium lead tri-iodide (MAPbI$_3$) being arguably the most
studied compound. Their success as light harvesters in solar cells
is due to the combined effect of their excellent absorption
properties \cite{Tanaka03}, a large carrier diffusion length
\cite{Stranks13,Xing13} and easy
fabrication \cite{burschka2013sequential,liu2013efficient}. This
striking success has generated research on a wide variety of
perovskite based photonic devices, including light emitting
diodes \cite{tan2014bright}, lasers \cite{zhu2015lead,Saliba15},
photodetectors \cite{fang2015highly}, and single photon
sources \cite{park2015room}.

A question currently under debate, important both from a fundamental point of view and for the optimization of perovskite based
devices, is the value of the exciton binding energy ($R^{*}$). Early magneto-optical studies \cite{Hirasawa94,ishihara1994optical,Tanaka03},
together with absorption measurements \cite{Incenzo14}, suggested an exciton binding energy $\simeq 40-50$\,meV for MAPbI$_{3}$.
This would imply that a significant fraction of photogenerated carriers form excitonic bound states at room temperature, with
important consequences for the architecture of solar cells. However, when the frequency dependence of the dielectric
constant \cite{Even14,Lin15} is taken into account, the estimated exciton binding energies are in the range
$2-15$\,meV \cite{Even14,Lin15,Yamada15,soufiani2015polaronic,Phuong2016}.
It has been suggested that discrepancies in the values reported for
the exciton binding energy might be due to the grain size and to the
environment surrounding the grains \cite{Incenzo14}. Polycrystalline
perovskites have a large density of defects at the surface and at
grain boundaries \cite{shkrob2014charge}, which influence the carrier
mobility and diffusion length, with an increased diffusion length in
films with enhanced grain size \cite{xiao2014solvent}. Another
suggestion is that the materials could exhibit microscopic
ferroelectric polarization, leading potentially to a dynamic Rashba
effect \cite{Zheng15, Etienne16} at high temperatures or to the
formation of nanoscale domains with random orientations at low
temperature \cite{Filipetti15}.  Any of these effects could result in
radical changes to device design becoming necessary depending on the
device production methods.
The recent synthesis of single crystal perovskites has allowed the
trap density to be reduced significantly, leading to a dramatically
enhanced diffusion length and carrier mobility
\cite{valverde2015intrinsic,dong2015electron,lian2016perovskite}.
The optical characterization of a single crystal methylammonium lead
tri-bromide suggests that the exciton binding energy is
$\simeq15$\,meV \cite{tilchin2016hydrogen} at low temperature,
however, in the absence of magnetic field, the exciton binding
energy could not be determined in the technologically important
cubic crystal phase at room temperature. A low exciton binding
energy would support the free carrier scenario at room temperature,
consistent with the photoluminescence of single crystal MAPbI$_{3}$
at 300 K \cite{fang2015photophysics}.

In this paper, we report on the first interband magneto-optical
studies of a single crystal hybrid organic-inorganic perovskite,
MAPbI$_{3}$, enabling the direct measurement of the exciton binding
energy and reduced mass in both the orthorhombic and tetragonal
phases. By introducing a derivative technique we demonstrate that
magneto reflectivity measurements can be analyzed as easily as
transmission and allow us to accurately determine these quantities
from which we can then deduce the dielectric constant. The values
obtained are identical to those measured in polycrystalline thin
films using magneto-absorption \cite{Miyata15}, unequivocally
demonstrating that the exciton binding energy has not been
influenced by the device processing procedures.

Single crystals of MAPbI$_3$ were grown from seed crystals in supersaturated solution and oriented by Laue diffraction. The
magnetoreflectivity measurements were performed in pulsed magnetic fields up to 66\,T, with a typical pulse duration of $\sim
300$\,ms. The sample was placed in a liquid helium cryostat and immersed in liquid or gaseous helium. White light, provided by a
broadband halogen lamp, was coupled to a multimode fiber. The reflected signal was collected by a fiber bundle in the Faraday
configuration with the light wavevector parallel to the applied magnetic field. A monochromator coupled to a liquid nitrogen
cooled CCD camera was used to analyze the reflected signal. Typical exposure time was 2\,ms, which allows spectra to be measured
at essentially constant magnetic field.

The reflectivity data were fitted using the relation
\begin{equation}\label{KKanalysis}
r(E)=r_{0}+ \sum_i A_{j} \Re \left(\frac{E_{j}-E+\text{i}\Gamma_{j}}{\Gamma_{j}^{2}+(E-E_{j})^{2}}\right)
\end{equation}
where $A_{j}, E_{j}$ and $\Gamma_{j}$ are the amplitude, energy and broadening parameter for each of the resonances
present \cite{Korona96}. $r_0$ is a constant (background), which controls the total amplitude of the reflectivity signal. In
practice taking the negative derivative of the fitted reflectivity with respect to energy, $(-\text{d}r/\text{d}E)$, produces peaks at almost
exactly the energies of the resonant absorption, as expected from a Kramers-Kronig analysis and shown in the  Supporting Information (S.I.). The data were therefore analyzed and could be more easily displayed by taking the negative derivative of the measured reflectivity.
\begin{figure}[t]
  \centering
    \includegraphics[width= 1.0\linewidth]{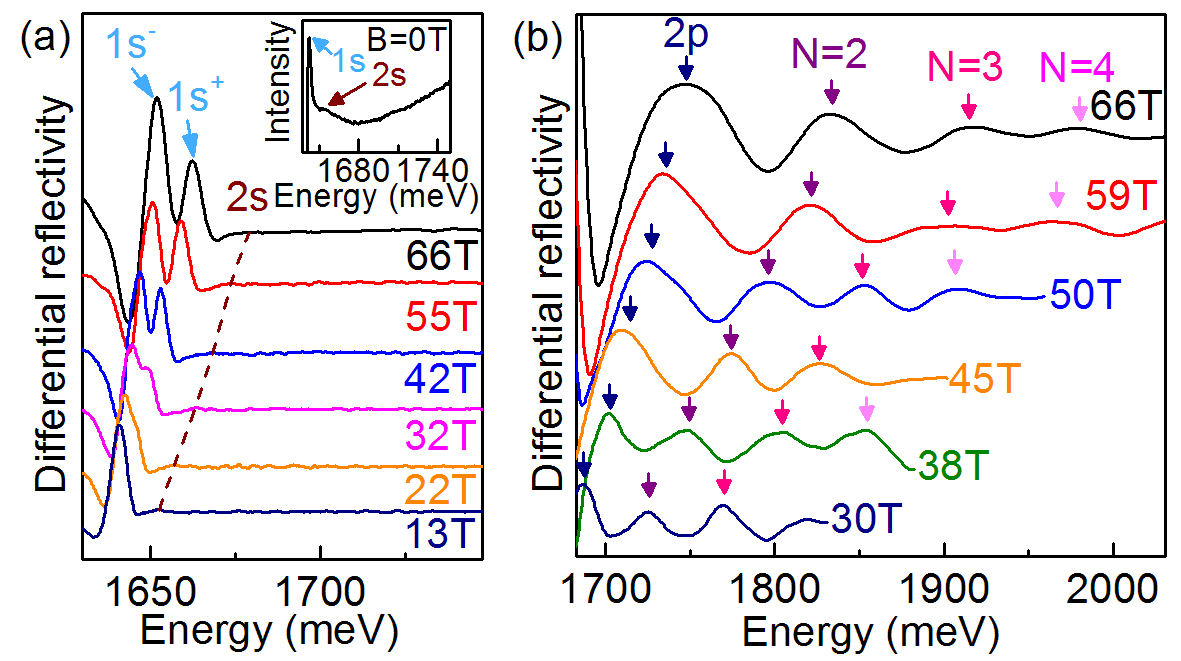}
  \caption{(a) Differential reflectivity
spectra measured at the indicated magnetic field values at 2K. The spectra are vertically offset for clarity. The arrows mark the
relevant absorption peaks. Inset: as-measured reflectivity spectrum at zero magnetic field. (b) High magnetic field differential
reflectivity spectra after normalization by zero field reflectivity. Arrows of the same colours indicate the transition between
the same Landau levels.}
  \label{fig:DifferentialReflectivity}
\end{figure}
Representative differential magneto reflectivity spectra measured at
2\,K are shown in Fig.\ \ref{fig:DifferentialReflectivity} for
magnetic fields up to 66\,T. In Fig.\
\ref{fig:DifferentialReflectivity}(a) we show the spectral region
where the strongest excitonic features are visible. At low magnetic
field, the pronounced peak corresponds to the absorption of the 1s
exciton state at an energy of 1640\,meV. The single crystal
MAPbI$_3$ shows a well resolved 1s transition which is fitted with a
broadening parameter of 3.8\,meV, compared with typical values of
$\gtrsim 20$\,meV observed for polycrystalline thin films
\cite{Miyata15}. This difference is attributed to the excellent
crystal quality even at the surface since the reflectivity is
dominated by the region within approximately one absorption length,
around 200nm, which is comparable to the thicknesses used for the
thin film samples. The narrow linewidth of the 1s transition enables
us to resolve an additional weak absorption peak on the high energy
side of the 1s peak, attributed to the 2s excitonic state, visible
here even at zero magnetic field. This is highlighted in the inset
of Fig.\ \ref{fig:DifferentialReflectivity}(a), where we show the
direct reflectivity at zero magnetic field $r(B=0)$.

Previously, in thin film MAPbI$_3$, the 2s state was only observed
as a weak shoulder at high magnetic fields \cite{Miyata15}, while
the large broadening of the 1s transition of Br-based MA perovskites
precluded the observation of the 2s state even at high magnetic
field \cite{galkowski2016determination}. The small linewidth of the
1s state also allows the observation of the Zeeman splitting of the
1s transition without the need for polarizing optics in the
detection path (the splitting is clearly seen in Fig.\
\ref{fig:DifferentialReflectivity}(a) for $B\geq 32$\,T). By
extracting the energies of the Zeeman-split 1s states from
magneto-reflectivity measurements, we obtained a linear splitting
yielding the effective $g$-factor for the Zeeman splitting
$g_\text{eff} = 2.66 \pm 0.1$. This value is slightly larger than
previously reported \cite{Tanaka03}, possibly due to the improved
resolution of our measurements, but nevertheless similar to our
previous results obtained on thin films
\cite{galkowski2016determination}. Finally, the high-energy
absorption peaks can be better seen by plotting the spectra obtained
by differentiating the spectrum normalized by the spectrum measured
at zero field. These spectra are shown in Fig.\
\ref{fig:DifferentialReflectivity}(b), where a series of maxima,
identified with the resonant absorption involving interband
transitions between Landau levels, can be seen.

The observation of two hydrogenic bound states accompanied by inter Landau level transitions gives a very well defined
measurement of some of the fundamental parameters for bulk MAPbI$_{3}$ such as the exciton binding energy and the reduced mass
which have proved to be highly
controversial \cite{Hirasawa94,ishihara1994optical,Tanaka03,Even14,Lin15,Yamada15,soufiani2015polaronic,Miyata15,galkowski2016determination}.
The measurements of these two parameters are effectively decoupled, as they influence the observed transition energies in
distinctly different regions of the energy spectrum. The energy $E_{n,0}(\gamma)$ of the hydrogen-like neutral exciton
transitions close to the band edge can be described using a numerical solution of the hydrogen atom in strong magnetic
field \cite{Makado19}. The transition energy depends on the dimensionless parameter $\gamma=\hbar \omega_{\text{c}} / 2R^*$, where
$\omega_{\text{c}}=\text{e}B/\mu$ is the cyclotron frequency, and $\mu$ is the reduced effective mass of the exciton, defined as
$\mu^{-1}=m_{\text{e}}^{-1} + m_{\text{h}}^{-1}$, where $m_{\text{e}}$ and $m_{\text{h}}$ denote the effective mass of the
electron and hole. The exciton binding energy is defined by
$R^*=R_0\mu/m_0\varepsilon_{\text{r}}^2$, where $R_0$ is the atomic
Rydberg, and $\varepsilon_{\text{r}}$ is the relative dielectric
constant of the material.

The Zeeman splitting of the excitonic transitions seen in Fig.
\ref{fig:DifferentialReflectivity}(a) is included by introducing a
Zeeman splitting term in the hydrogen-like absorption spectrum
\cite{galkowski2016determination}. Additionally, at high magnetic
fields where $\gamma>1$ (the fitted parameters yield $\gamma>1$
above 28T) the higher energy excitonic transitions approach the free
carrier interband transitions between Landau levels
\cite{Watanabe03}, with energies given by
\begin{equation}
E_{n}^{\text{LL}}(B) = E_{\text{g}} + \left(n+\frac{1}{2} \right)\hbar\omega_{\text{c}} \pm \frac{1}{2}g_{\text{eff}} \mu_B B, \label{eq:LL}
\end{equation}
where $E_{\text{g}}$ is the band gap, $n=0, 1, 2, \dots$ represents the Landau orbital quantum number in the valence and
conduction bands and $\mu_B$ is the Bohr magneton.

\begin{figure}[t]
\centering
\includegraphics[width= 1.0\linewidth]{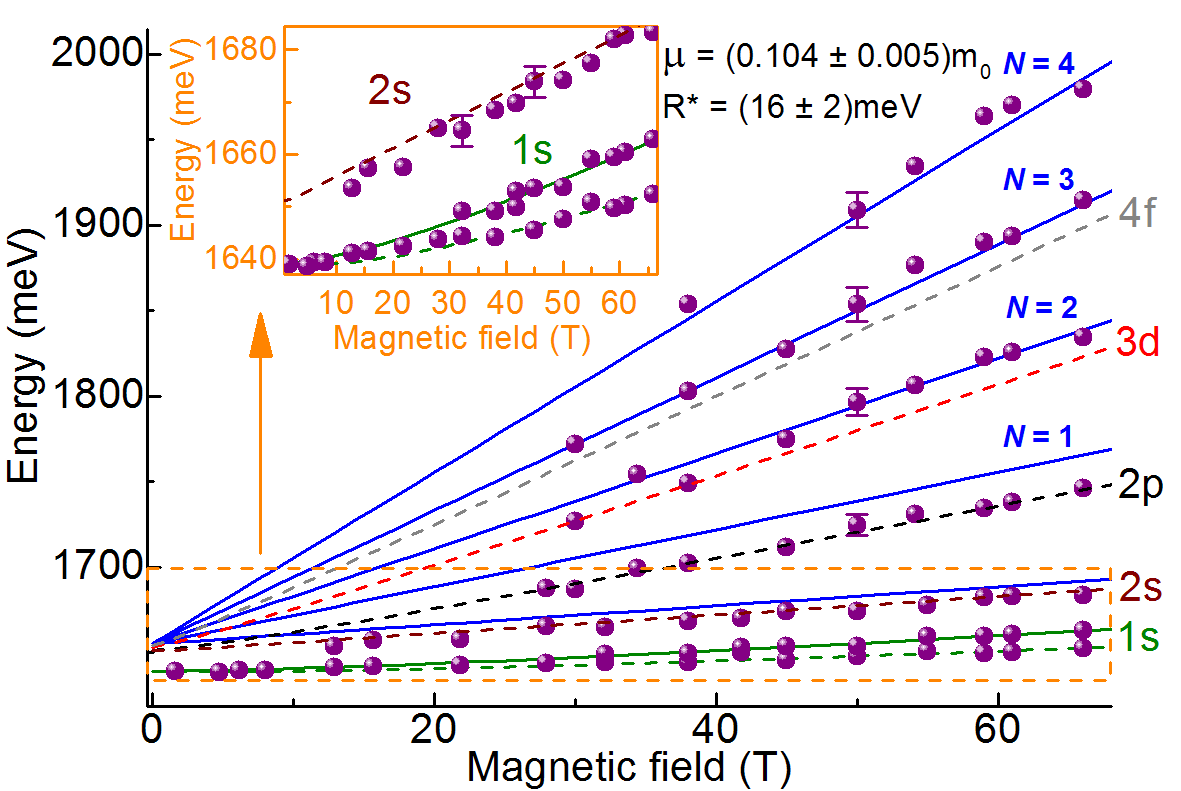}
\caption{Energies for excitonic and Landau level transitions as a function of the applied magnetic field at 2K. Blue,
solid lines are the results of linear fits of the interband transition between Landau levels in the valence and conduction bands.
Dashed lines result from the fit of hydrogen-like transitions. Inset: close-up view of the low energy transitions.}
\label{fig:Fanchart}
\end{figure}

In Fig.\ \ref{fig:Fanchart}, we show the full set of observed
transitions. The higher energy dipole allowed ($\Delta n = 0$)
interband Landau level transitions are fitted with Eq.\,\ref{eq:LL}.
The dominant fitting parameter is the reduced effective mass of the
exciton, which is determined to be $\mu = (0.104 \pm 0.005) m_0$,
where $m_0$ is the free electron mass. This is consistent with the
theoretically calculated values from density functional theory of
$\mu=0.099m_0-0.11 m_0$,\cite{Menendez14,Umari14} and in excellent
agreement with the experimentally determined effective mass for a
300\,nm thick polycrystalline film deposited on a glass substrate
\cite{Miyata15}using similar techniques to those used for typical
devices \cite{Stranks15}. The reduced mass is then used as a fixed
parameter in the fitting of the excitonic transitions, which
strongly constrains the value of the exciton binding energy, which
must also be consistent with the 1s-2s separation observed at zero
field. The zero field splitting follows the series of 3 dimensional
hydrogen-like energy states
\begin{equation}
E_{i}^{\text{ex}}=E_{\text{g}} - \frac{R^*}{i^2},
\label{eq:EnergyHydrogen}
\end{equation}
where $E_{i}^{\text{ex}}$ is the energy of the $i^{\text{th}}$ excitonic level. The fitting allows us to conclude that at
2\,K,  $R^*=16\pm 2$\,meV. This value is significantly smaller than
early estimates at low temperature based on magneto-optical
measurements of only the 1s
state \cite{Hirasawa94,ishihara1994optical,Tanaka03} or temperature
dependent absorption measurements \cite{Incenzo14} (37-50\,meV), but
agrees well with recent experimental results on thin
films \cite{soufiani2015polaronic,Miyata15,Lin15,Even14,galkowski2016determination,Yamada15}.
In particular the observation of the precise 1s - 2s separation at
zero magnetic field provides further verification of the
magneto-optical extrapolation analysis used by previous works where
the 2s state could only be observed at high magnetic fields and
numerical fitting of the 2s field dependence was
required \cite{Miyata15,galkowski2016determination}.

 The early results depended entirely on the correct estimation of the dielectric constant, which was taken to be close to the high frequency value, on the assumption that a reduced dielectric screening occurs when the exciton binding energy is larger than the optical phonon energy \cite{Hirasawa94}. The phonon structure is however considerably more complex \cite{Lin15}, and the observation of optical phonon modes with energies from 8\,meV to 16\,meV \cite{quarti2013raman,Phuong2016}, suggests enhanced  dielectric screening and, consequently, a value intermediate between the static $\varepsilon_0=25.7$ \cite{Wehrenfennig14}, and the high frequency
 $\varepsilon_{\infty}=5.6$\, \cite{Umari14,Brivio14} dielectric constant should be used \cite{Even14}.

The observation of higher exciton energy levels and Landau states enables us to measure the exciton binding energy without making
any assumptions about the dielectric constant, which we can then deduce directly from the above definition of the effective
Rydberg to be $\varepsilon_{\text{r}}\sim9.4$ \cite{Miyata15}, which is intermediate between the low and high frequency values given above, as expected.  This also allows us to deduce an effective exciton Bohr radius, $a_{\text{B}}^{*}$ of 4.6 nm. The
exciton binding energy obtained for the bulk MAPbI$_{3}$ crystal is thus identical, within experimental accuracy, to the binding
energy of a randomly oriented polycrystalline film, determined with the same fitting procedure \cite{Miyata15}. Our results imply
that contrary to some suggestions \cite{Incenzo14}, both the effective mass and the exciton binding energy are largely independent of the crystallinity and crystal orientation
of the sample, as would be expected since the value of $a_{\text{B}}^{*}$ is substantially larger than the size of the polycrystalline
grains in typical device structures and the band structure is predicted to be essentially isotropic \cite{Menendez14,Umari14}.

\begin{figure}[t]
  \centering
  \includegraphics[width= 1.0\linewidth]{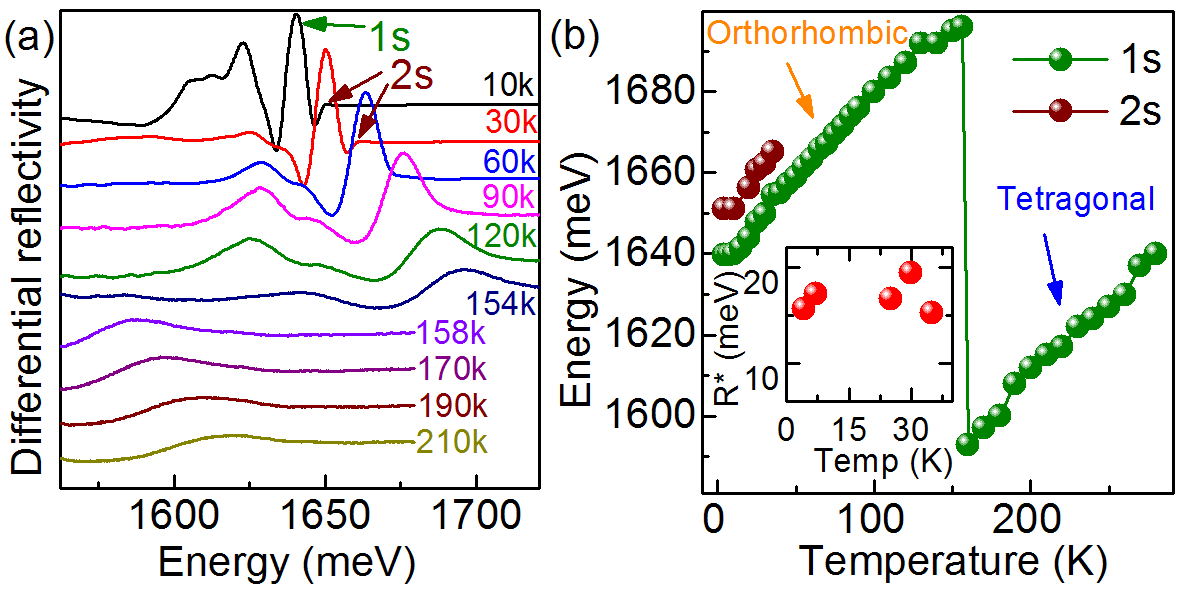}
  \caption{(a) Differential reflectivity spectra measured at different temperatures. (b) 1s (green circles) and 2s
(red circles) energy as a function of temperature. The inset indicates the exciton biding energy (estimated here from the 1s-2s
splitting) as a function of the temperature.}
  \label{fig:Tdependence}
\end{figure}

Practical device applications of perovskites, however, require a thorough understanding of the high temperature behavior of these
materials, in particular in the high temperature tetragonal crystal phase which occurs above approximately
150\,K.\cite{baikie2013synthesis} The phase transition from orthorhombic to tetragonal is accompanied by a significant change of
bandgap ($\sim 100$\,meV)\cite{Incenzo14,Yamada15,Even14}, as can be seen in Fig.\ \ref{fig:Tdependence}(a) which shows
differential reflectivity spectra at different temperatures. At zero magnetic field, we observe both 1s and 2s transitions at
temperatures as high as 35\,K, due to the excellent crystalline quality. The 1s and 2s peaks exhibit a similar blue shift with
increasing temperature, as shown in Fig.\ \ref{fig:Tdependence}(b), following the trend already observed for the 1s state in thin
film samples \cite{Incenzo14,Yamada15,Miyata15}. From 1s-2s separation we deduce R$^{*}$ is independent of the temperature up to
35K as shown in the inset in Fig.\,\ref{fig:Tdependence}(b). Above 154\,K there is an abrupt transition to a lower energy peak,
red shifted by 103\,meV with respect to the orthorhombic phase, which suggests that the entire area sampled by the excitation
spot has completed the transition to the tetragonal phase.

\begin{figure}[t]
  \centering
  \includegraphics[width= 1.0\linewidth]{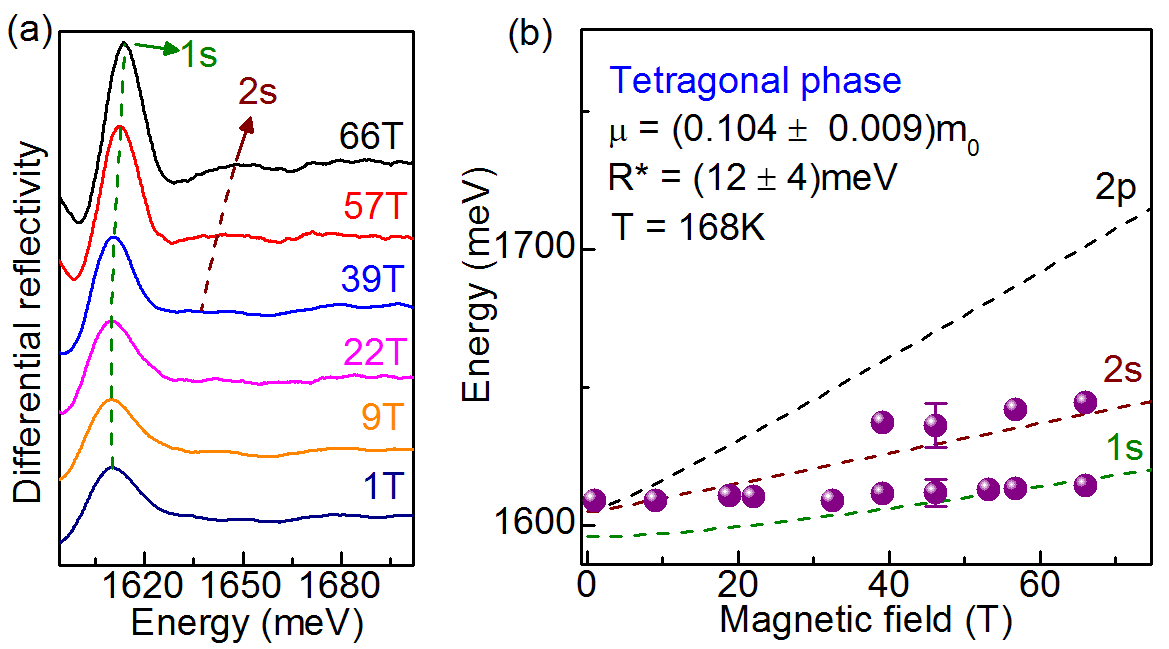}
  \caption{(a) Magneto
reflectivity spectra measured at $T=168$\,K differentiated with
respect to energy. (b) High temperature energy fan chart. The dashed
lines represent hydrogenic transitions.}
  \label{fig:HighTfanChart}
\end{figure}

We have repeated the magneto reflectivity measurements at 168\,K and differential reflectivity spectra are shown in Fig.\
\ref{fig:HighTfanChart}(a). The 1s state appears as a pronounced peak in the reflectivity spectrum even at low magnetic field,
but with a considerable broadening (FWHM of 22\,meV). A shallow peak related to the absorption of the 2s state can be identified
for magnetic fields larger than 39\,T (Fig.\ \ref{fig:HighTfanChart}(a)) but no higher energy transitions could be reliably
identified. Figure \ref{fig:HighTfanChart}(b) shows the transition energy fan chart at 168\,K where the reduced mass has been
assumed to be the same as at low temperature, $\mu \simeq 0.104 m_0$. The exciton binding energy is expected to be more strongly
affected by the phase transition. In the tetragonal phase, an increase of the dielectric constant is expected, due to the dynamic
disorder related to the rotational motion of the organic cation enabled by the structural transition \cite{Poglitsch87}. An
increase of the dielectric screening should lead to a reduction of the exciton binding energy, with values reported for room
temperature ranging from 5 to 12\,meV \cite{Even14,Yamada15,soufiani2015polaronic}.

We have fitted the data for the 1s and 2s transitions above 30\,T with the procedure described above and find a high field exciton binding
energy of $12 \pm 4$\,meV, comparable to that estimated by performing similar high temperature magneto transmission experiments
on a thin film \cite{Miyata15}. We consider this value to be an upper bound for the exciton binding energy at zero field, as the
high frequency cyclotron motion in the high magnetic fields is expected to reduce the effective dielectric screening and hence
increase the exciton binding energy at high fields. We estimate that reasonable bounds for the high temperature value of $R*$ are in the region of 5-12 meV, which suggests that the high temperature dielectric constant is in the range of $\varepsilon_{\text{r}}\sim11-17$, consistent with the increased screening due to the rotational motion of the organic cations discussed above \cite{Even14}.

In conclusion, we have studied the magneto optical properties of high quality single crystal MAPbI$_3$, which exhibits narrow 1s
and 2s excitonic lines at low temperature and at zero magnetic field. Detailed magneto reflectivity measurements reveal that the
exciton binding energy and effective masses are identical to those extracted from similar measurements on device quality polycrystalline thin
films deposited on glass substrates. The measured exciton binding energies, 16\,meV for the low temperature orthorhombic phase
and $\lesssim 12$\,meV for the room temperature tetragonal phase, are lower than the thermal energy at room
temperature, consistent with organic-inorganic semiconducting perovskites showing non-excitonic behavior in solar cells and other
devices.

\begin{acknowledgement}

This work was partially supported by ANR JCJC project milliPICS, the
R\'egion Midi-Pyr\'en\'ees under contract MESR 1305303, STCU project
5809, the BLAPHENE project under IDEX program Emergence and NEXT
ANR-10-LABX-0037 in the framework of the ``Programme des
Investissements d'Avenir''. This work was supported by EPSRC (UK) via
its membership to the EMFL (grant no.\ EP/N01085X/1). Z.Y. acknowledges
financial support from the China Scholarship council. A.A.H.
acknowledges the support of the EPSRC Platform Grant (Grant No.\
EP/M020517/1).

\end{acknowledgement}

\section{Supporting Information Available}

Supporting information:Crystal growth procedure, X-ray
characterization of single crystal MAPbI$_3$, representative fits of
differential reflectivity spectra, temperature dependence of
linewidth of differential reflectivity.


\bibliography{BibPerovskite}

\end{document}